\documentclass[submission,copyright,creativecommons]{eptcs}
\providecommand{\event}{GASCom 2024} 

\usepackage{iftex}

\ifpdf
  \usepackage{underscore}         
  \usepackage[T1]{fontenc}        
\else
  \usepackage{breakurl}           
\fi

\usepackage{graphicx} 
\graphicspath{{Images/}}
\usepackage{subfig}
\usepackage{wrapfig}
\usepackage{amssymb}
\usepackage{amsmath}
\usepackage{hyperref}
\usepackage[english]{babel}

\title{Uniform Sampling and Visualization of 3D Reluctant Walks}
\author{
Benjamin Buckley \qquad\qquad Marni Mishna\thanks{This work is generously funded by NSERC Discovery Grant (Canada) RGPIN-04157}
\institute{Simon Fraser University\\Burnaby, BC, Canada}
\email{\quad bbuckley@sfu.ca \quad\qquad mmishna@sfu.ca}
}
\def\titlerunning{Uniform Sampling and Visualization of 3D Reluctant Walks}
\def\authorrunning{B. Buckley \& M. Mishna}
\begin{document}
\maketitle

\begin{abstract}
A family of walks confined to the first orthant whose defining stepset has drift outside of the region can be challenging to sample uniformly at random for large lengths. We address this by generalizing the 2D walk sampler of Lumbroso et al. to handle 3D walks restricted to the first orthant. The sampler includes a visualizer and means to animate the walks.
\end{abstract}

\begin{wrapfigure}{R}{0.45\textwidth}
    \centering
    \includegraphics[width=0.70\linewidth]{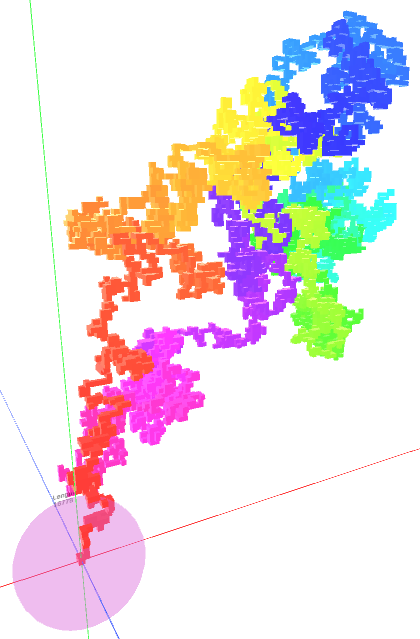}
    \caption{\small \emph{A walk of length 16,778 generated uniformly at random from a stepset with drift (-1, -1 -1). Progression of the walk is given by hue, with the initial position at $(0,0,0)$ in red and the final position in magenta.}}
    \label{fig:longwalk}\vspace{-20pt} 
\end{wrapfigure}
\section{Introduction} A combinatorial class of lattice walks is a set of walks, where each step is taken from a finite set of possibilities,  and that the walks remain within some region (typically a cone centered at the origin).  The simplicity of lattice walks contributes to their ubiquity. Indeed, many families of discrete objects have natural bijections to lattice walks models. Visualizations of the large scale behaviour of uniformly generated random walks can reveal underlying structure of the walks, and consequently, of related objects. 

Here, we will focus on the uniform random generation of 3-D walks confined to the first orthant ($\mathbb{Z}_{\geq 0}^3$), in particular models that are said to be \emph{reluctant} because the drift of the stepset (the vector sum) is outside of the cone. The random sampling algorithm naturally extends the 2D sampling algorithm for reluctant walks of Lumbroso et al.~\cite{lumbroso_taming_2017}. Fig.~\ref{fig:longwalk} illustrates a visualization of a 16,778 step reluctant walk restricted to the first orthant, (with drift $(-1, -1, -1)$) taken from the Unity interface we have developed. We have represented the walk steps using blocks, and have coloured the walk in a way to make its evolution clear.

\subsection{Motivation}
When the stepset has zero drift, naive generation (where steps are drawn at random, and a walk is rejected once it leaves the region) can be effective. However, when the drift is outside of the region, particularly when it fits the criterion (defined below) of being reluctant such a strategy is defeated by the very small proportion of unrestricted walks that remain in the region. 

Some recent asymptotic studies of weighted walk models have provided explicit examples of complete asymptotic formulas in terms of the weights and have pointed to some potential sources of interesting phenomena. For example in both~\cite{courtiel_weighted_2017}, and \cite{mishna_asymptotics_2020} the authors fix a stepset, and then  determine asymptotic counting formulas as a function of weightings of each allowable step. In both cases, the exponential growth factor in the dominant term is smooth as a function of the weights, but the sub-exponential growth is defined piecewise in terms of the drift of the model. That is to say, as the drift changes the asymptotic growth formula undergoes phase transitions as the drift crosses into different regions. One of our motivations is to understand how these changes influence a uniformly chosen model near these critical drift regimes. A robust, easy to use visualizer can  be useful to identify phenomena that may subsequently be established analytically or algebraically.

\subsection{Definitions}
To describe the algorithm, we first make some notation explicit. We focus on finite stepsets $\mathcal{S}\subset \mathbb{Z}^3$ with the property that  the defining set of vectors are not contained in any half-space. A walk starts at the origin, takes steps in $\mathcal{S}$ such that there is no step outside of the first orthant, $\mathbb{N}^3$. Although our random sampler is generic, in this abstract we present various integer weightings of $\{\pm e_1, \pm e_2, \pm e_3\}$. In this context a model is reluctant if the drift has all negative components, that is $\sum_{s\in\mathcal{S}} s \in \mathbb{Z}_{<0}^3$.  We also consider weightings that result in a drift outside the first orthant.

\section{Algorithm}
The sampling algorithm we use, like the 2D sampling algorithm of Lumbroso \emph{et al.}~\cite{lumbroso_taming_2017} has three key elements.  
It is a rejection algorithm, but rather than generating an unrestricted walk, our algorithm generates walks confined to a judiciously chosen half-space containing the first orthant, and rejects generated walks if they exit the first orthant. The set of walks confined to a half space is in bijection with a set of generalized Dyck words, for which a grammar can be made explicit~\cite{duchon_enumeration_2000}. Given the grammar, uniform sampling can be effectively accomplished using a Boltzmann sampler. We visualize the walks in Unity to give access to an interface, and open the possibility to animation. Images and videos demonstrating this are available to view at \url{https://benbuckleyanimator.wixsite.com/portfolio/randwalkviz}. 

\subsection{A well chosen half-space}
From our restriction on $\mathcal{S}$ we can deduce that there exists a half space (and possibly many) so that, asymptotically,  the number of walks confined to that half space has the same exponential growth factor as walks further confined to the first orthant. We say more about this in the example below, but generically this follows from a result of Garbit and Raschel~\cite{garbit_exit_2016}. This means that the proportion of the walks the half-space that are \emph{not} in the first orthant grows sub-exponentially with length. 

Since walks confined to a half space only interact with a single boundary, these walks are in bijection with a one dimensional model. To define this one dimensional model, the stepset $\mathcal{S}$ is projected onto a vector $\mathbf{v}=(a,b,1) \in \mathbb{R}^3$ orthogonal to the hyperplane defining the halfspace to create a 1D stepset. We can make this set explicit: $\mathcal{A}_\mathbf{v}=\{ai+bj+k\mid (i,j,k)\in\mathcal{S}\}$, with steps in $\mathbb{R}$.
The exponential growth of the asymptotic number of walks in the first orthant model is the equal to the minimum value of the Laurent polynomial $S(x,y,z)=\sum_{(i,j,k)\in \mathcal{S}}x^iy^jz^k$ attained in a suitable domain of $\mathbb{R}_{>0}^3$. Suppose the minimum value is attained at $(x^*,y^*,z^*)$. Provided $z^*\neq 1$,  we can show that the 1D model associated to $\mathbf{v}=(a,b,1)$ with $a=\log(x^*)/\log(z^*)$, $b=\log(y^*)/\log(z^*)$ is such that the exponential growth of the asymptotic number of walks in the model is the same as the orthant model. Remark, that because of our conditions on $\mathcal{S}$ we have $x^*y^*z^*\neq 0$. In the case where $z^* = 1$, we can project on to $(a,b,0)$ with $a=\log(x^*)$ and $b=\log(y^*)$. If  $x^*=y^*=z^*=1$, then any halfspace containing the orthant can be used.

\subsection{Generalized Dyck words}
The previous process yields a one-dimensional stepset with real valued steps. In order to create a grammar for a Boltzmann sampler, we need a stepset with integer valued steps. We find a rational approximation -- that is, a stepset with steps in $\mathbb{Q}$ sharing a small denominator, typically $< 10$ for ease of computation -- and then multiply through by the denominators to obtain a stepset with steps in $\mathbb{Z}$. The closer the approximation, the less the rejection although we will always generate walks in the orthant uniformly. In our example, the previous step yields integer steps directly. 

Given a finite set of integers,  the grammar for 1D walks that do not go below 0 is described in ~\cite{duchon_enumeration_2000}. We have applied to this to our running example in Fig.~\ref{fig:example_grammar} below.

\subsection{Boltzmann generation}
Given the grammar, the sampling is done using a Boltzmann sampling strategy, following~\cite{duchon_boltzmann_2004}, although any random generating strategy for a grammar could be used in its place. 
Here we describe only how we compute some of the required elements. 

Given the grammar, we used Maple's Combstruct library to solve the algebraic equations to obtain explicit generating function for these 1-dimensional walks. It turns out that this is the most difficult step for computational reasons: when stepset has large integers, the number of rules in the grammar can grow quickly, potentially inhibiting the computation of the generating function. 

When an explicit generating function is not determined a high order series approximation is used in the computation.

Given the generating function, we determine its dominant singularity, and in the case of a series approximation we determine an approximation. Boltzmann sampling uses generating function evaluations at, or near, this singularity.

Boltzmann sampling can generate objects of any size, but, for a given size the sampling is uniform. To generate large objects, one uses evaluation points close to or at the dominant singularity, provided it is not a pole. In these problems, the singularities are branch points, not poles, and so we can evaluate directly at the dominant singularities. 

\subsection{Running the algorithm}
Thus, in summary, for a given model $\mathcal{S}$, we first have a set up phase to  first determine the best half-space, then generate the corresponding grammar, and do the computational preamble (generating function evaluations) for the Boltzmann sampler. To generated a walk, we run the sampler, biject back to a 3D walk, and eliminate those walks that leave the first orthant. The preamble is done in Maple, but we run the sampler in Unity. 

\section{Drawn walks}
\subsection{The reluctant model in Fig.\ref{fig:longwalk}}
\label{sec:reluctant}
The walk in Fig.~\ref{fig:longwalk} is sampled from the set of walks with the stepset $\{ e_1, e_2, e_3, -e_1, -e_1, -e_2, -e_2, -e_3, -e_3\}$. Note that this is a multiset with each of the negative steps appearing twice, increasing their probability. We calculate the drift by taking the vector sum of these steps, and obtaining $-e_1 -e_2 - e_3 = (-1,-1,-1)$. We note that this model fits in the framework of Theorem 2 in \cite{mishna_asymptotics_2020}, from which we deduce that the number of walks of length $n$ grows, up to a constant,  like $(6\sqrt{2})^n\,n^{-3}$. Thus, the exponential growth factor is $6 \sqrt{2}$.

\paragraph{Find the halfspace}
The inventory of the stepset is $S(x,y,z) = \sum_{(i,j,k) \in \mathcal{S}} x^i y^j z^k = x+2/x+y+2/y+z+2/z$. A gradient computation shows this function is minimized at $\sqrt{2}(1,1,1)$. The optimal vector is  $\mathbf{v}=(a,b,1)$ with $a = \frac{\log(x^*)}{\log(z^*)}$ and $b = \frac{\log(y^*)}{\log(z^*)}$, that is, $\mathbf{v}=(1,1,1)$. The stepset $\mathcal{S}$ is projected onto this vector to obtain a 1-dimensional stepset with steps in $\mathbb{R}$:
\begin{gather*}
\mathcal{A}_{\mathbf{v}} = \{1, 1, 1, -1, -1, -1, -1, -1, -1\}. 
\end{gather*}
Remark that the inventory of this walk is $A(u)=3u+6/u$. Because the drift is negative, by Banderier and Flajolet~\cite[Theorem 4]{banderier_basic_2002}, the exponential growth factor for the walks that are never negative is $A(\tau)$, where $\tau$ is the critical point of $A(u)$. In this case $A'(\sqrt{2})=0$ and $A(\tau)=2\sqrt{6}$. Furthermore, we know that the number of walks of length $n$, grows like $(2\sqrt{6})^n n^{-3/2}$ (up to a multiplicative constant) We are assured: This walk has an identical exponential growth factor to the 3D walks. The proportion of these walks that are images of walks remaining in the first orthant is approximately $\frac{n^{-3}}{n^{-3/2}}= n^{-3/2}$, for length $n$.

\paragraph{Find the grammar}

As the steps are all integer valued, we can proceed directly to the grammar phase. 
We use the formulation of grammars for generalized Dyck words, as described in \cite{duchon_enumeration_2000}, to obtain a grammar for half-hyperplane walks in our 1-dimensional stepset. Following this specification, we obtain the grammar in Figure \ref{fig:example_grammar}. 
\begin{figure}\small
\begin{eqnarray*}
\mathcal{P} &=& \mathcal{D} \times \mathcal{P}_{aux} \\
\mathcal{P}_{aux} &=& \epsilon + \mathcal{L}_1 \times \mathcal{P}_{aux} \\
\mathcal{L}_1 &=& a_1 \times \mathcal{D} + a_2 \times \mathcal{D} + a_3 \times \mathcal{D} \\
\mathcal{R}_1 &=&  b_1 \times \mathcal{D} + b_2 \times \mathcal{D} + b_3 \times \mathcal{D} + b_4 \times \mathcal{D} + b_5 \times \mathcal{D} + b_6 \times \mathcal{D}  \\
\mathcal{D} &=& \mathcal{L}_1 \times \mathcal{R}_1 + \epsilon \\
a_1,...,a_3 &=& \text{Atoms representing $+1$ steps} \\
b_1,...,b_6 &=& \text{Atoms representing $-1$ steps}
\end{eqnarray*}
\caption{\emph{\small Grammar for 1-dimensional walks with inventory $A(u)=3u+6/u$}}
\label{fig:example_grammar}
\end{figure}

\paragraph{Boltzmann sampling}
To generate a walk, we next set up the Boltzmann sampler. This requires generating functions for each non-terminal in the grammar. The generator requires an evaluation point, which we take as the dominant singularity $\frac{1}{6\sqrt{2}}$. Walks that are deemed too short or outside the target length are rejected. The inverse bijection is applied to determine the corresponding 3-dimensional step from which it was projected. The walk is accepted if it stays in the orthant, and rejected if it exits.

\paragraph{Verification} Since there is some computational approximation in the Boltzmann sampler, we did some tests to convince ourselves that we were achieving uniform generation. We generated 10,000 walks of length 10, and tallied the number of walks ending at each point. The proportions were compared with the exact proportion of walks expected to end at each point, calculated using recurrences. As the number of walks increases, we calculate the root-mean-square error between the discrete distribution generated by our results, and the distribution implied by exact counting. The root-mean-square error decreases from $\approx 0.00340361673$ when generating 10 walks, $\approx 2.815461678 \times 10^{-6}$ when generating 1,000,000 walks. While this isn't a proof, it does lend confidence that the algorithm generates walks uniformly.

\begin{figure}
\centering
\vspace{-5mm}
\hfill\subfloat[Drift $(0,0,0)$, approximately 1000 steps]{
  \includegraphics[width=.4\textwidth]{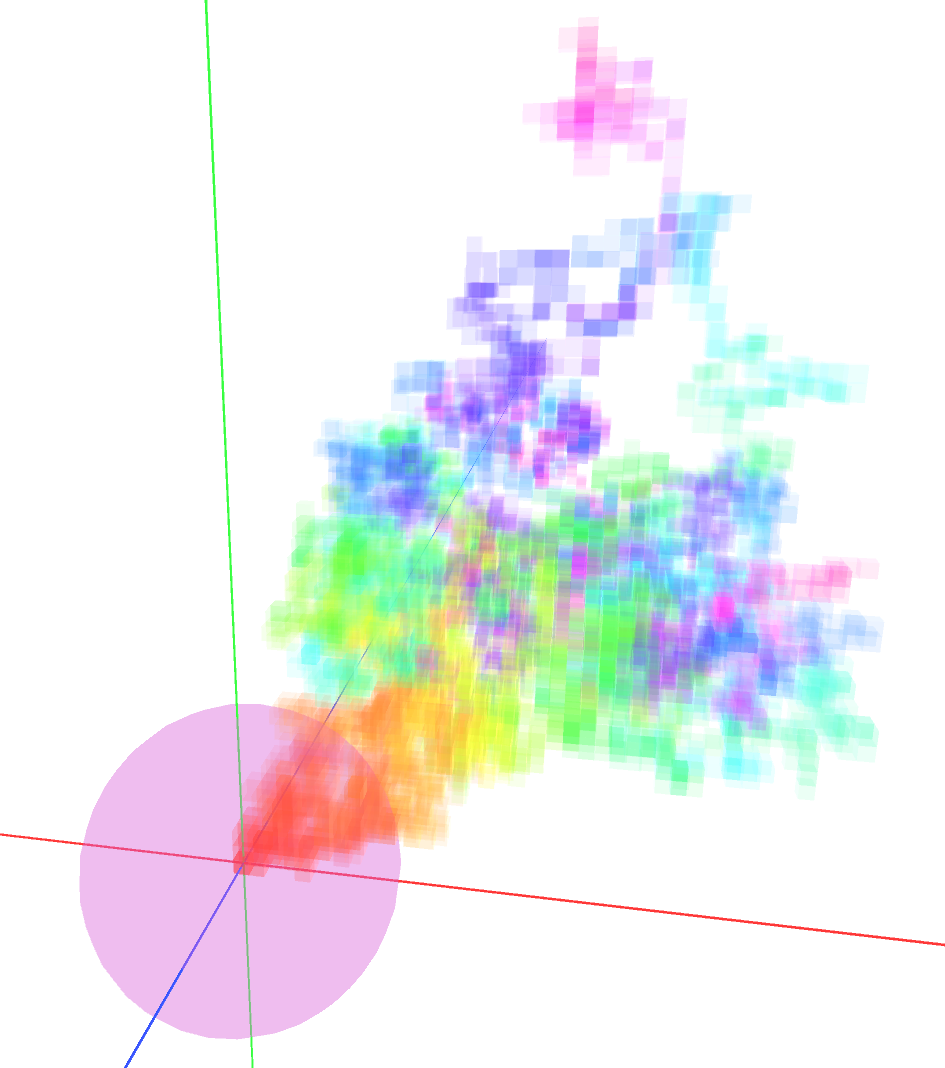}
}
\hfill
\subfloat[Drift $(-1,-1,-1)$, approximately 1000 steps]{
  \includegraphics[width=.4\textwidth]{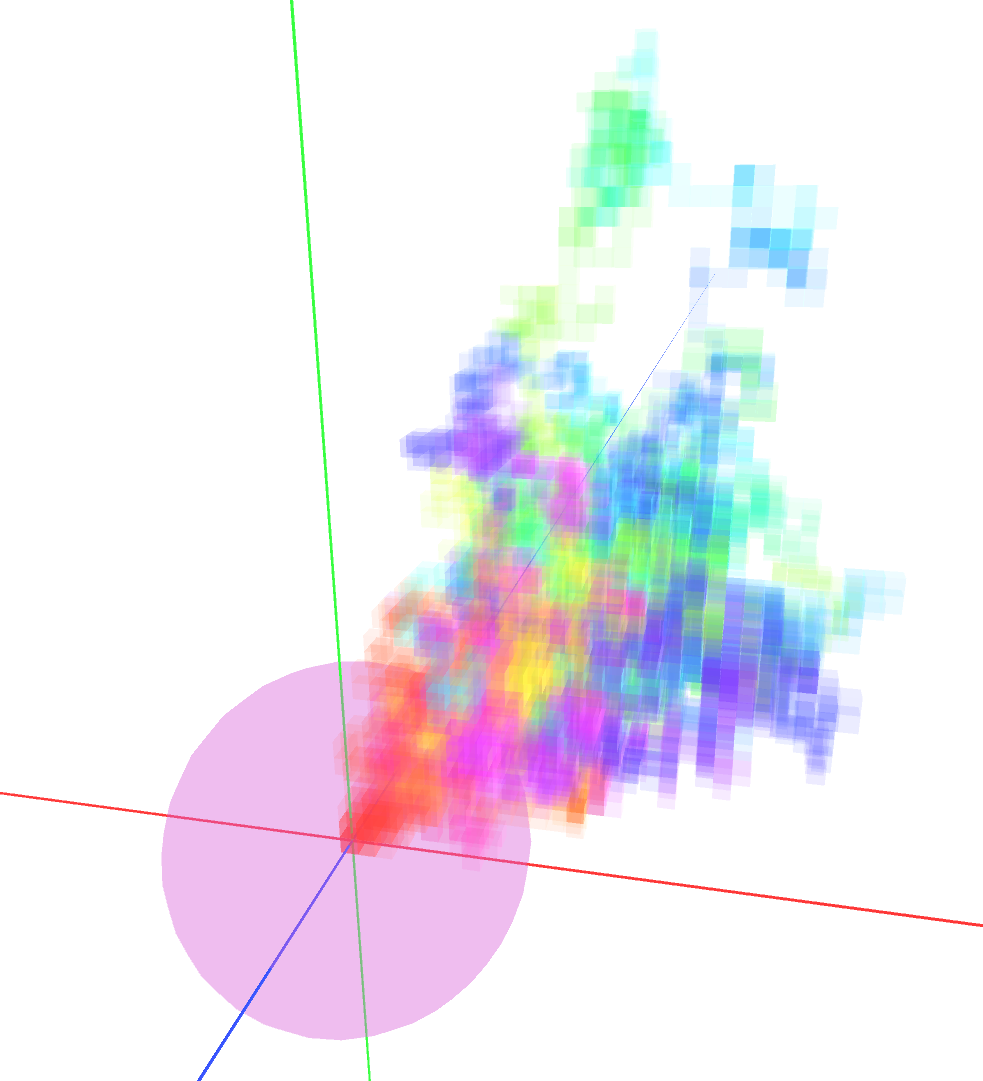}
}\hfill\mbox{}\\
\subfloat[Drift $(0,-1,0)$, approximately 1000 steps]{
  \includegraphics[width=.4\textwidth]{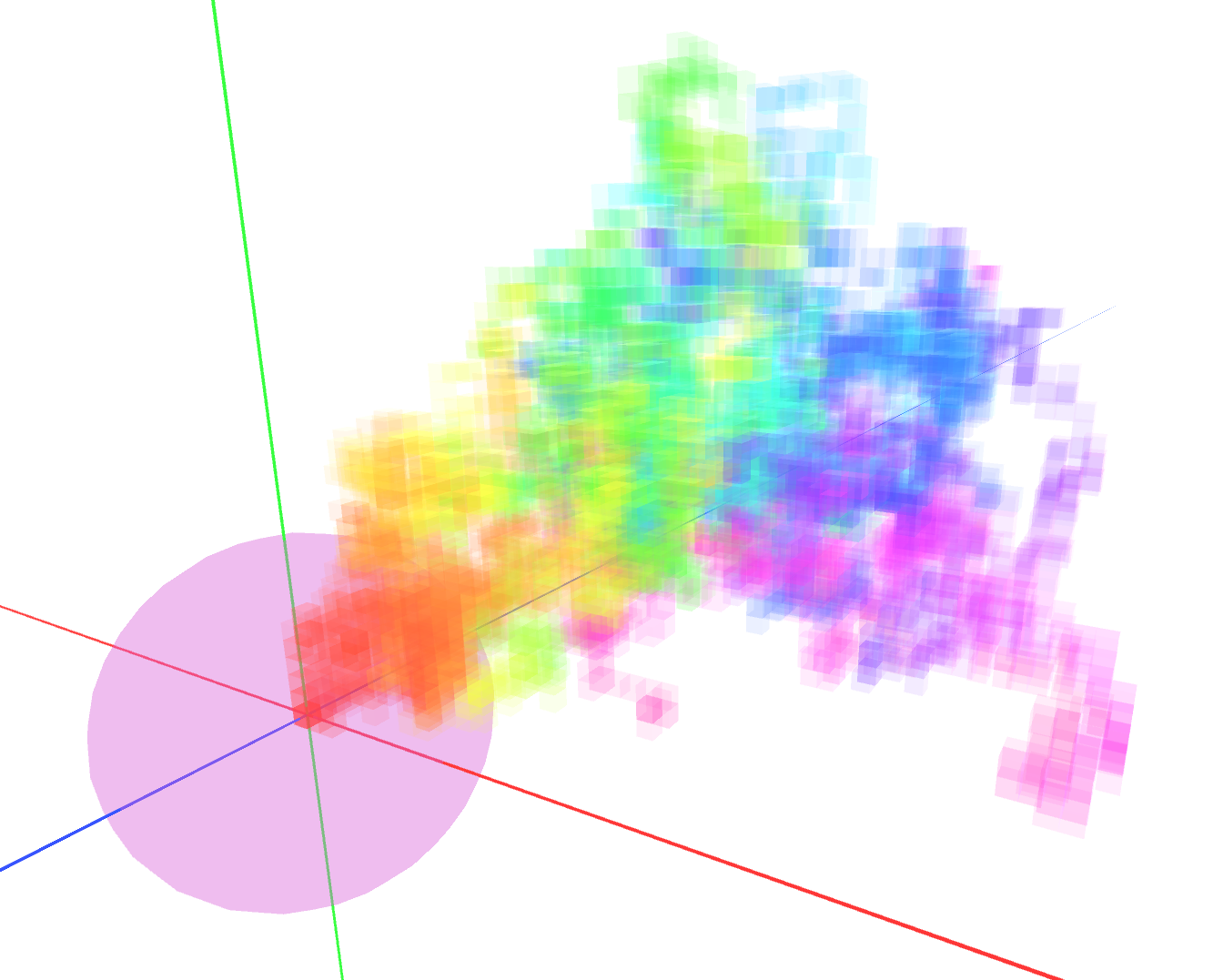}
}\hfill
\hfill\subfloat[Drift $(0,-1,-1)$, approximately 1000 steps]{
  \includegraphics[width=.4\textwidth]{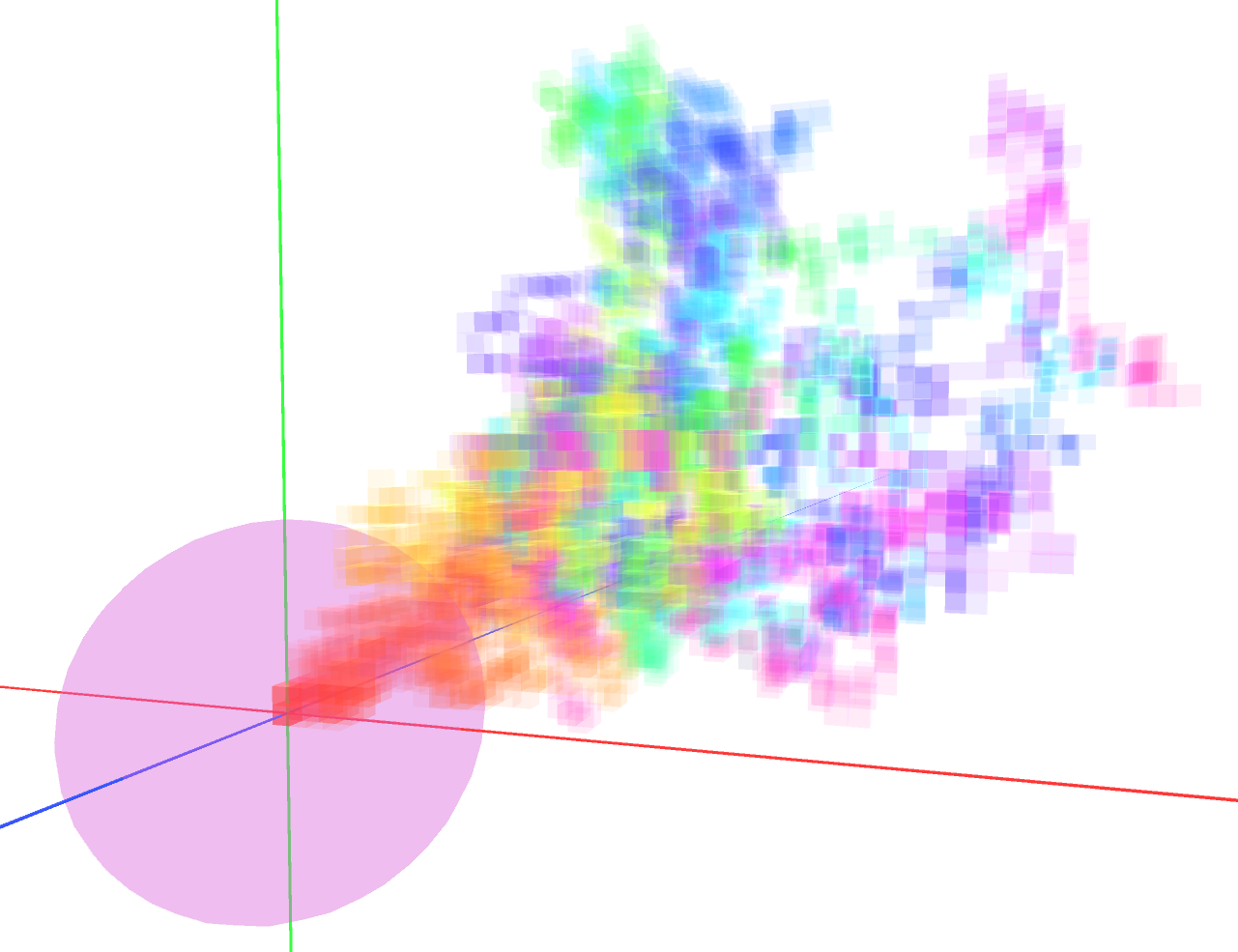}
}\hfill\mbox{}
\caption{\emph{\small Randomly generated lattice walks with various stepsets. Images each show 10 walks in the same space. A translucent sphere with radius 20 is centered at the origin to give a sense of scale.}}
\label{four-walks}
\end{figure}
\subsection{Comparative study of models}
Some examples of walks in $\mathbb{Z}^3$, restricted to the positive octant, can be seen in Fig.~\ref{four-walks}. Multiple walks are shown in each image.

The development of a typical walk changes significantly depending on the drift.

For example, note that in Fig.~\ref{four-walks}(b), when the drift tends towards the XZ plane, the walks tend to form an arc from the origin towards a point further out on the XZ plane. In contrast, in Fig.~\ref{four-walks}(d), then the drift tends towards the origin, a cluster of magenta near the origin and green on the outer regions suggests that walks tend to venture far away from the origin, not returning until the end of the walk.

Our algorithm makes it possible to generate long reluctant walks with high probability in cases where a generating a walk of comparable length  under the naive scheme is virtually statistically impossible. Using the approximate number of orthant walks of length $n$ to be $(6\sqrt{2})^nn^{-3}$, we can see that in the naive scheme a walk of length $n$ has probability on the order of $\left(\frac{6\sqrt{2}}{9}\right)^n n^{-3}$ of being generated in a given run. This is less than .0002\% when $n=100$, where as the Boltzmann sampler has probability $n^{-3/2}$, which is $0.1\%$ for $n=100$. Both take strategies take linear time to generate a walk of length $n$.

To be even more concrete, we generated 10 walks with drift $(-1,-1,-1)$ of length around 100 using both strategies.  With the naive rejection 133,065,405 walks were generated in total in order to find 10 that remained in the positive octant, the rest being rejected. This took approximately 8 minutes and 8.7 seconds. In contrast, when using Boltzmann sampling targeting lengths between 95 and 105 steps, it was only necessary to generate 202,669 walks to find 10 that were the appropriate length and remained in the positive octant. This was done in approximately 8.4 seconds. Both computations were done on similar conditions on a 2020 Macbook Pro with an Apple M1 chip.

\begin{wrapfigure}{hR}{0.35\textwidth}
\centering
\vspace{-7mm}
  \includegraphics[width=40mm]{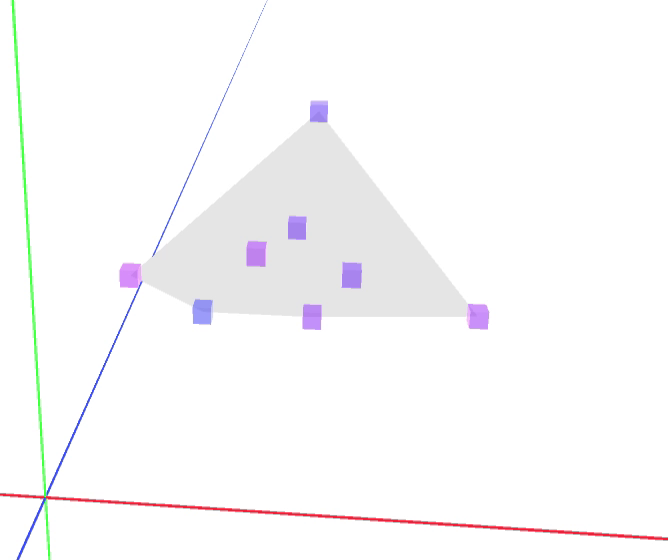}
\caption{{\small\emph{The convex hull at step 220 in the progression of 10 walks.}}}
\label{convexhulls}

\end{wrapfigure}

\section{Future work}

One of the appeals of this project -- and one of the reasons we've created it in the game engine Unity -- is that the application provides the user with multiple ways to visualize and understand walks. The reader is encouraged to visit the site for some animations associated with reluctant walks. 

We have discussed the colour coding already. It is also possible to animate sets of walks, with each frame showing a simultaneous position of every walk in the set. Recently, we've made it possible to view the convex hull of the simultaneous positions of the walks at a given step, with the convex hull changing as the walk progresses. (See Figure \ref{convexhulls}).

Other future plans include tools for generating data to give a sense of how quickly the walks are generated with different methods, and how many rejections occur. We also plan to create animations visualizing walks in terms of their bijection with other objects -- for example, queues and tableaux. We also hope to make the Unity interface available in the future.
\newpage
\nocite{*}
\bibliographystyle{eptcs}
\bibliography{generic}
\end{document}